\documentclass[aps,prb,reprint,amsmath,superscriptaddress,amssymb,graphicx]{revtex4-1}
\usepackage{amsfonts}
\usepackage{bbm}
\usepackage{graphicx}
\usepackage{dcolumn}
\usepackage{bm}
\usepackage{subfigure}
\usepackage{amssymb}
\newcommand{\me}{\mathrm{e}} \newcommand{\mi}{\mathrm{i}}\newcommand{\piup}{\mathrm{\pi}}
\newcommand{\dif}{\mathop{}\!\mathrm{d}}
\DeclareMathAlphabet{\mathfsl}{OT1}{cmss}{m}{sl}

\begin{document}
	\title{Quantum transport through 3D topological insulator PN junction under magnetic fields}
	\author{Ning Dai}	
	\author{Yan-Feng Zhou}
	\author{Peng Lv}
	\affiliation{International Center for Quantum Materials, School of Physics, Peking University, Beijing 100871, China}
	\author{Qing-Feng Sun}
	\email{sunqf@pku.edu.cn}
	\affiliation{International Center for Quantum Materials, School of Physics, Peking University, Beijing 100871, China}
	\affiliation{Collaborative Innovation Center of Quantum Matter, Beijing 100871, China}
	\affiliation{CAS Center for Excellence in Topological Quantum Computation, University of Chinese Academy of Sciences, Beijing 100190, China}
	\date{\today}

\begin{abstract}
The 3D topological insulator (TI) PN junction under magnetic fields presents a novel transport property which is investigated both theoretically and numerically in this paper.
Transport in this device can be tuned by the axial magnetic field.
Specifically, the scattering coefficients between incoming and outgoing modes oscillate with axial magnetic flux at the harmonic form.
In the condition of horizontal mirror symmetry, the initial phase of the harmonic oscillation is dependent on the parities of incoming and outgoing modes.
This symmetry is broken when a vertical bias is applied which leads to a kinetic phase shift added to the initial phase.
On the other hand, the amplitude of oscillation is suppressed by the surface disorder while it has no influence on the phase of oscillation.
Furthermore, with the help of the vertical bias, a special (1,-2) 3D TI PN junction can be achieved, leading to a novel spin precession phenomenon.
\end{abstract}

\maketitle

\section{Introduction}

Topological insulator (TI) is a sort of electronic materials
which have insulating bulk states and conducting surface states
protected by time-reversal symmetry.\cite{TI1,TI2,TI3,TI4}
This conception was first proposed in 2D materials\cite{HgTe} and then
generalized into 3D materials.\cite{predict1,predict2,predict3}
The experimental realization of 3D TI was reported in several materials
such as Bi$_{1-x}$Sb$_x$,\cite{BiSb1} Bi$_2$Te$_3$,\cite{BiTe1} Sb$_2$Te$_3$\cite{LL4SbTe1}
and Bi$_2$Se$_3$.\cite{BiSe1,BiSe2}
With strong spin orbital coupling in 3D TI, the spin direction of surface state
is constrained perpendicular to the momentum,\cite{modeth,modet}
which is called spin-momentum locking and can be observed by angle-resolve photoemission spectroscopy (ARPES).\cite{mode0}
Due to this spin-momentum locking property, an electron acquires $\piup$ Berry phase
by going around the surface.\cite{ar1,zyf,berry1,berry2,berry3,berry4}
This phase can be compensated by a magnetic flux, leading an Aharonov-Bohm (AB) oscillation
in conductance which has been observed in a number of experiments.\cite{AB1,AB2,AB3,AB4}
With a strong magnetic field added perpendicular to the upper and lower surfaces,
the energy spectrum of the upper and lower surfaces are reorganized
into Landau-Level (LL),\cite{LL1,LL3,ar2,LJYin}
and the chiral modes emerge along the side surfaces\cite{ar2}.

Tuning the potential on the two adjoint regions of a 3D TI ribbon
by gate voltage\cite{gate1,gate2,gate3} or chemical doping\cite{doping1},
the 3D TI ribbon is turned into a 3D TI PN junction.
Under perpendicular magnetic field, chiral modes come into being at the PN boundary and side surfaces.\cite{3dtipnj}
Based on the chiral modes, in 2015, a spin-based Mach-Zehnder interferometry has
been proposed on this 3D TI PN junction with a strong vertical magnetic field and an axial magnetic flux.\cite{core}
The conductance in this device can be tuned with the magnetic flux which presents
a harmonic oscillation. In addition, the spin polarization of the reflected and transmitted currents
are opposite, which enables this device to work as a spin splitter.

In this paper, we take a thorough investigation in the transport property of 3D TI PN junction under magnetic field.
The transport property is described by the scattering coefficients between the incoming and outgoing modes which are calculated by the Green's function method.
These scattering coefficients, as well as the conduction, oscillate with magnetic flux at the harmonic form.
The initial phase of the oscillation is 0 or $\piup$ dependent on the parities of modes in the pristine 3D TI PN junction where the horizontal mirror symmetry is conserved and the parity is well defined.
The symmetry is broken when a non-zero vertical bias is applied.
This vertical bias changes the potential energy in both upper and lower surfaces, leading to a kinetic phase shift in the oscillation.
In some cases, the vertical bias changes the number of modes and generates a special (1,-2) 3D TI PN junction, in which a novel spin precession phenomenon happens.
We also investigate the effect of disorder on the lower surface in imitation of substrate.
The phase shift of oscillation at every single disorder configuration is neutralized when it averages in a large configuration ensemble.
Instead, the averaged conductance oscillation shows a suppression in amplitude.

The rest of this paper is organized as follow:
In Sec.\ref{model} we give the form of Hamiltonian and illustrate the modes
which constitute the transport path.
In Sec.\ref{coefficient} we introduce the scattering coefficient and
give an overview of the conductance in different conditions.
Sec.\ref{parity} shows the parity effect on scattering coefficients in symmetric 3D TI PN junction. Sec.\ref{bias} shows the two effects of vertical bias: kinetic phase (Sec.\ref{phase})
and spin precession (Sec.\ref{precession}).
Sec.\ref{disorder} presents the effect of disorder on the lower surface of 3D TI PN junction.
Finally, we conclude the work in Sec.\ref{conclusion}.

\section{\label{model}model and modes}

\begin{figure}
	\includegraphics[width=\linewidth]{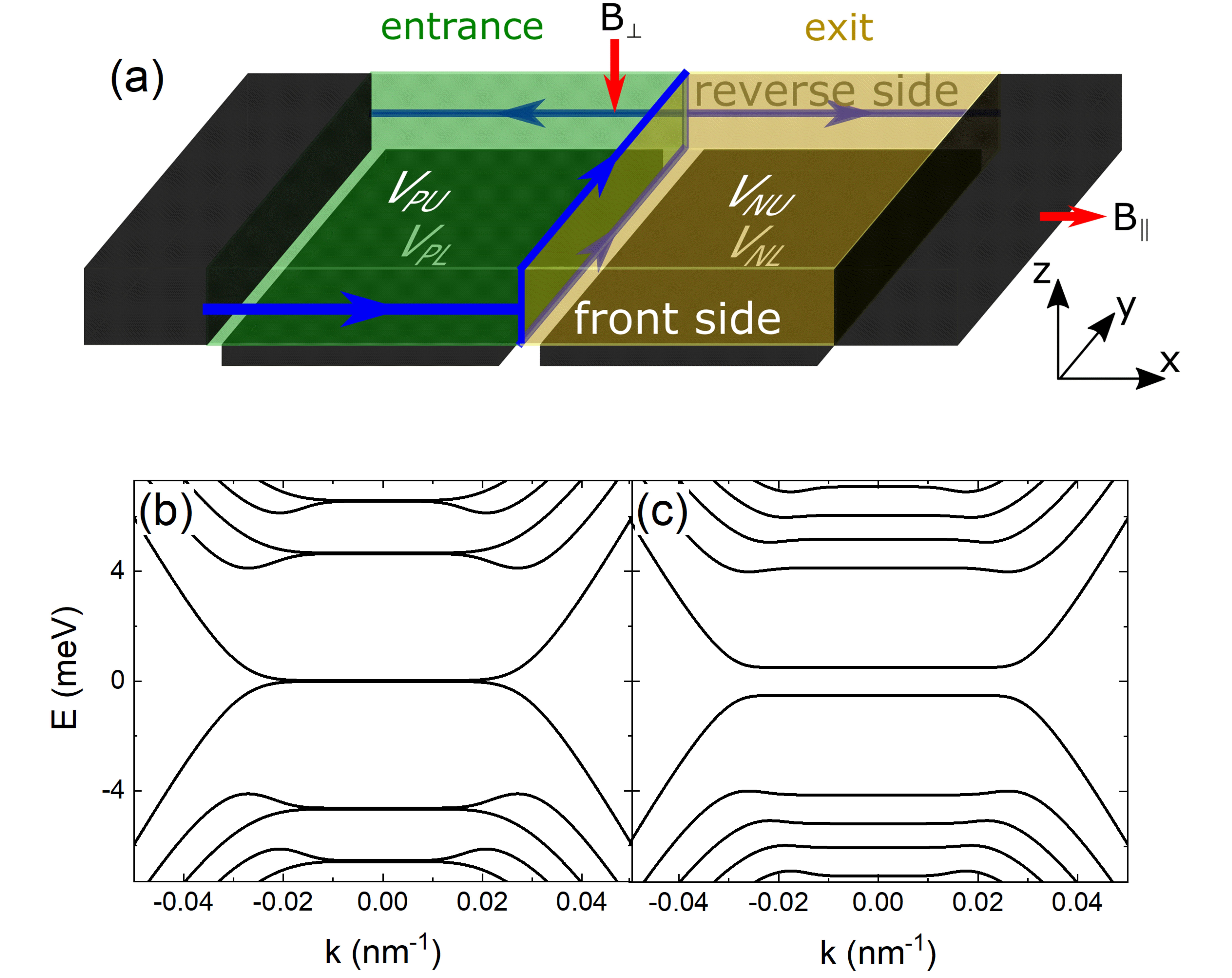}
 \caption{\label{fig1}
 (a) One scheme of 3D TI PN junction device.
 Green part represents the entrance region while yellow represents the exit. The vertical bias is the potential difference between the upper and lower surfaces,  $U_T=V_{PU}-V_{PL}=V_{NU}-V_{NL}$.
  (b) and (c) are the energy spectrum of 3D TI in a perpendicular magnetic field
 with the vertical bias $U_T=0$ (b) and $U_T=1$meV (c).
 }
\end{figure}

One of the scheme of 3D TI PN junction is shown in Fig.\ref{fig1}(a), which is a square nanoribbon.
Cartesian coordinate system is established of which the x axis is along the injecting direction
of electrons and z is along the normal vector of the upper and lower surfaces.
The original point is in the center of this 3D TI nanoribbon.
The nanoribbon is divided into two regions by y-z plane:
the entrance region in which electrons are injected, and the exit region.
The potential of entrance region is tuned by gate voltage $V_P$, while $V_N$ tunes the exit region.
The perpendicular magnetic field $B_\perp$ is along the -z direction.
Axial magnetic field $B_\parallel$ generates a flux $\phi=eLlB_\parallel/\hbar$
where $L$ and $l$ denote the length of y and z direction respectively.
If the 3D TI PN junction is made by chemical doping,
the upper and lower surfaces might be symmetric.
However, in the scheme presented in Fig.\ref{fig1}(a), the gates generate a perpendicular
electrostatic field and lead to a vertical bias between upper and lower surfaces.
Without loss of generality, we introduce a bias term $U_T$ to take this possible bias
into consideration.

We adopt a lattice model to describe the surface of 3D TI PN junction.
Although there exists a fermion doubling problem in the lattice model of 3D TI,\cite{double}
this trouble can be solved by introducing a Wilson term in the Hamiltonian.\cite{zyf}
The discretized lattice Hamiltonian of the surface of 3D TI PN junction is\cite{zyf}
\begin{eqnarray}
		H &=&
  \sum\limits_{u=1}^N\sum\limits_vc^\dagger_{uv}(T_0+V_{uv})c_{uv}
		+\sum\limits_{u=1}^N\sum\limits_v \me^{\mi\Phi_u}c^\dagger_{u,v+1}T_1c_{uv} \nonumber\\
& +& \sum\limits_v \left[\sum\limits_{u=1}^{N-1}\me^{\mi\frac{\phi}{N}}c^\dagger_{u+1,v}T_2c_{uv}
  -\me^{\mi\frac{\phi}{N}}c^\dagger_{1v}T_2c_{Nv}\right] +\mathbf{H.c.}	\label{ham}
\end{eqnarray}
where
\begin{eqnarray}
		T_0 &=&4\frac{W}{a}\sigma_z ,\nonumber\\
		T_1 &=&\mi\frac{\hbar v_f}{2a}\sigma_y-\frac{W}{a}\sigma_z ,\nonumber\\
		T_2 &=&-\mi\frac{\hbar v_f}{2a}\sigma_x-\frac{W}{a}\sigma_z .
\end{eqnarray}
The surface of the 3D TI PN nanoribbon is discretized into infinite layers along the $x$ direction
and each layer is discretized into $N$ sites.
The sites are labeled by $(u,v)$, where $u$ and $v$ are integer
with $v$ ($v\in (-\infty,\infty)$) being the layer index and
and $u$ ($u\in [1,N]$) being the index
along the circumference of the 3D TI ribbon.
$c_{uv}^\dagger$ and $c_{uv}$ creates and annihilates an electron on site $(u,v)$.
$\Phi_u$ and $\phi/N$ are the AB phases generated by perpendicular and axial magnetic fields, respectively.
$\Phi_u=B_\perp y_u$ where $y_u$ denotes the y coordinate of site $u$.
In the following numerical calculation, $\phi$ is in units of $\phi_0=\hbar/2e$, and we also take
the lattice constant $a=10$nm and the Fermi velocity $v_f=5\times10^5$m/s.
This Fermi velocity $v_f$ corresponds to topological insulator $\mathrm{Bi}_2\mathrm{Te}_3$
and $(\mathrm{Bi}_{x}\mathrm{Sb}_{1-x})_2\mathrm{Te}_3$,\cite{vfadd1,vfadd2}
which can be experimentally extracted from ARPES
and scanning tunneling spectroscopy (STS).
The Wilson term factor $W=0.06\hbar v_f$ which is brought in artificially to avoid fermion doubling problem.
Without loss of generality,
the electrostatic potential $V_{uv}$ is a combination of $V_P$,$V_N$ and $U_T$:
$V_{uv} = V_P+\frac{U_T}{2}\frac{z_u}{l}$ in the entrance region and
$V_{uv} = V_N+\frac{U_T}{2}\frac{z_u}{l}$ in the exit region,
where $z_u$ is the z coordinate of site $u$.
In particular, the potentials on the upper entrance, lower entrance, upper exit, and lower exit surfaces
respectively are $V_{PU}=V_P+U_T/2$, $V_{PL}=V_P-U_T/2$, $V_{NU}=V_N+U_T/2$, and $V_{NL}=V_N-U_T/2$,
with $\nu_{PU}$, $\nu_{PL}$, $\nu_{NU}$, and $\nu_{NL}$ being the corresponding filling factors.
Then $\nu_P = \nu_{PU}+\nu_{PL}$ and $\nu_N =\nu_{NU}+\nu_{NL}$ denote the filling factors
in entrance ($P$) and exit ($N$) regions.

In this paper, we study the quantum transport through 3D TI PN junction, which requires
the dimensions of the device smaller than the coherence length.
As for 3D TI, the coherent length can be measured by Aharonov-Bohm oscillation.
In Ref.[\onlinecite{AB3}], it gives the coherent length $L_\varphi$
to be $L_\varphi>0.5\mu m$ at 2K and $L_\varphi>2\mu m$ at 50mK.
For this reason, we focus on the PN junction with its width being hundreds of nanometers.
The quantum transport path in 3D TI PN junction consists of two different kinds of modes,
namely, longitudinal modes on side surfaces and transverse modes along the PN interface
on upper and lower surfaces.
Fig.\ref{fig1}(a) sketchily illustrates the transporting route of an electron along the modes.
Incident electrons along the longitudinal mode on the front side surface arrive
at the PN interface and then split into two branches of transverse modes
going along the upper and lower PN boundaries respectively.
The two branches converge at the reverse side surface with different phase
which determines whether it will be scattered into forward or backward outgoing mode.

The longitudinal modes are the eigenstates of the 3D TI perfect nanoribbon.
For a 3D TI perfect nanoribbon, there are four surfaces, i.e. the upper, lower,
front side and reverse side surfaces. Here the upper and lower surfaces are
much wider than the two side surfaces [see Fig.\ref{fig1}(a)].
So the surfaces of the 3D TI ribbon can be viewed as two homogeneous surfaces
(upper and lower surfaces) coupling by the narrow side surfaces.
Away from the edges, the electrons in upper and lower surfaces are localized
by the perpendicular magnetic field, corresponding to the double degenerate flat LLs
in the middle of Fig.\ref{fig1}(b).
At the edges, the eigenstates in the upper and lower surfaces are coupled
through the side surfaces, splitting into two branches in energy spectrum,
which is in analogy with the bonding and anti-bonding states in molecule.
The two branches derived from one LL hold different parity
under horizontal mirror reflection [see Fig.\ref{fig4}(a)].
This parity has an effect on the scattering coefficients which will be detailed in Sec.\ref{parity}.
The parity is broken under a nonzero vertical bias $U_T$
where each degenerate LL in upper and lower surfaces split into two LLs
at a distance of $U_T$ [Fig.\ref{fig1}(c)].

\begin{figure}
	\includegraphics[width=\linewidth]{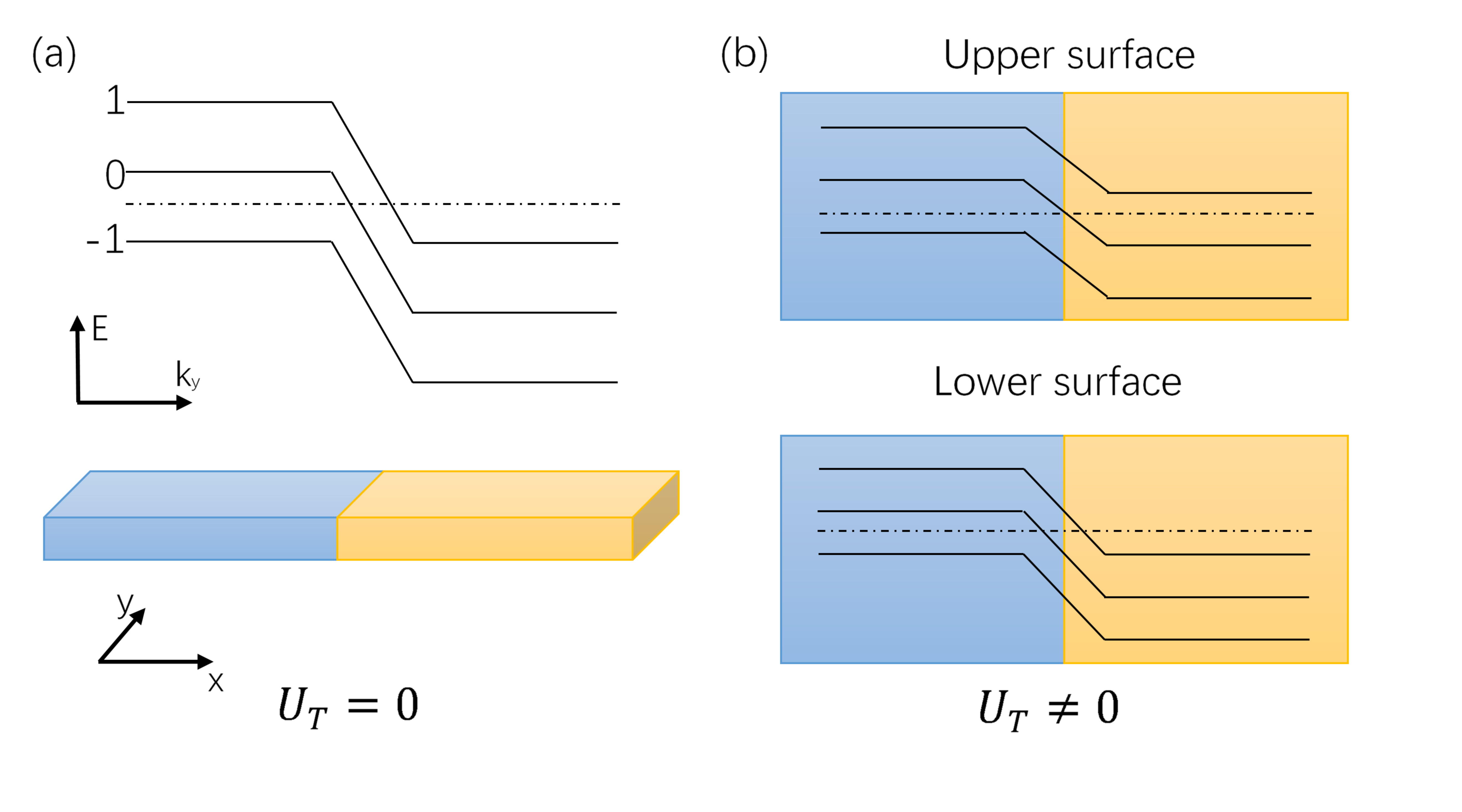}
	\caption{\label{fig2} A sketch of spectrum for transverse modes in 3D TI PN junction.
(a) The energy spectrum of transverse modes in upper and lower surfaces.
LLs deform in the vicinity of PN interface due to the
the potential difference in entrance and exit regions.
Some LLs cross the Fermi level and form the transverse modes.
(b) The vertical bias breaks the symmetry of upper and lower surfaces,
and may induce different number of transverse modes in the upper and lower surfaces.}
\end{figure}

On the other hand, the transverse mode is derived from LLs
in upper and lower surfaces [Fig.\ref{fig2}(a)].
On the upper or lower surface of the 3D TI, the LLs are flat.
However, these flat LLs are deformed near the PN interface,
raising at one region and descending at the other region.
If one LL crosses the Fermi level during this process, one transverse mode appears.
Analytically, the number of transverse modes on the upper and lower surfaces respectively are
$n_U=|\nu_{PU}-\nu_{NU}|$ and $n_L=|\nu_{PL}-\nu_{NL}|$.
In the absence of vertical bias ($U_T=0$),
the upper and lower surfaces are symmetric and their transverse modes are same also ($n_U=n_L$).
But when a nonzero vertical bias applied, this symmetry is broken and
this may induce different number of transverse modes in some situations [Fig.\ref{fig2}(b)].

Consider a 3D TI PN junction with the filling factor being $\nu_P$ in entrance region
and $\nu_N$ in exit region [hereafter denoted as $(\nu_P,\nu_N)$ for simplicity].
There are $|\nu_P|$ incident longitudinal modes $|\mathrm{in}\rangle_m$ on the front side surface,
and $|\nu_P|+|\nu_N|$ outgoing longitudinal modes $|\mathrm{out}\rangle_n$ on the reverse side surface,
$n=1,\cdots,|\nu_P|$ denoting backward modes in entrance region
and $n=|\nu_P|+1,\cdots,|\nu_P|+|\nu_N|$ denoting forward modes in exit region.
The transverse modes on the upper and lower surfaces are denoted by $|\mathrm{up}\rangle_i$ and $|\mathrm{low}\rangle_j$ respectively, where $i=1,\cdots,n_U$ and $j=1,\cdots,n_L$.

\section{\label{coefficient} conductance of 3D TI PN junction}

In general, every incoming mode $|\mathrm{in}\rangle_m$ goes through the path
in Fig.\ref{fig1}(a) and is scattered into outgoing modes $|\mathrm{out}\rangle_n$
both backward and forward.
We use the scattering coefficient $|S_{nm}|^2$ to describe the possibility
that an electron in mode $|\mathrm{in}\rangle_m$ is scattered into mode $|\mathrm{out}\rangle_n$,
namely,
\begin{equation}
|S_{nm}|^2=|\langle\mathrm{out}_n|S|\mathrm{in}\rangle_m|^2 .
\end{equation}
The conservation of probability ensures that $\sum_n |S_{nm}|^2 =\sum_m |S_{nm}|^2=1$.
The conductance $G$ of 3D TI PN junction is determined by the modes scattered forward,
namely,
\begin{equation}
G=\frac{e^2}{h}\sum_m\sum\limits_{n=\nu_P+1}^{\nu_P+\nu_N}|S_{nm}|^2 . \label{conduct}
\end{equation}
For the simplest case in a (1,-1) 3D TI PN junction where $\nu_P=-\nu_N=1$,
the conductance has been proven harmonic oscillating with the axial magnetic flux $\phi$.\cite{core}
The situation for multimodes is similar, since $|S_{nm}|^2$ presents
a same oscillation in the absence of $U_T$ (Sec.\ref{parity}).
The vertical bias $U_T$ induces a kinetic phase shift in scattering coefficient oscillation,
which is discussed in Sec.\ref{phase}.

Based on the lattice Hamiltonian Eq.(\ref{ham}),
the Green's function method is applied
to solve the scattering amplitude $S_{nm}$
as well as conductance $G$ in 3D TI PN junction.\cite{my}
The detailed procedure (formula) for solving $S_{nm}$ can refer Ref.[\onlinecite{my}].
The Green's function method is flexible and available in different condition
such as large filling factor, vertical bias, and disorder.
In the numerical calculation, we set the perpendicular magnetic field $B_\perp=659$Gs,
the TI nanoribbon height $l=100$nm and width $L=700$nm, if not mentioned intentionally.

\begin{figure}
	\includegraphics[width=\linewidth]{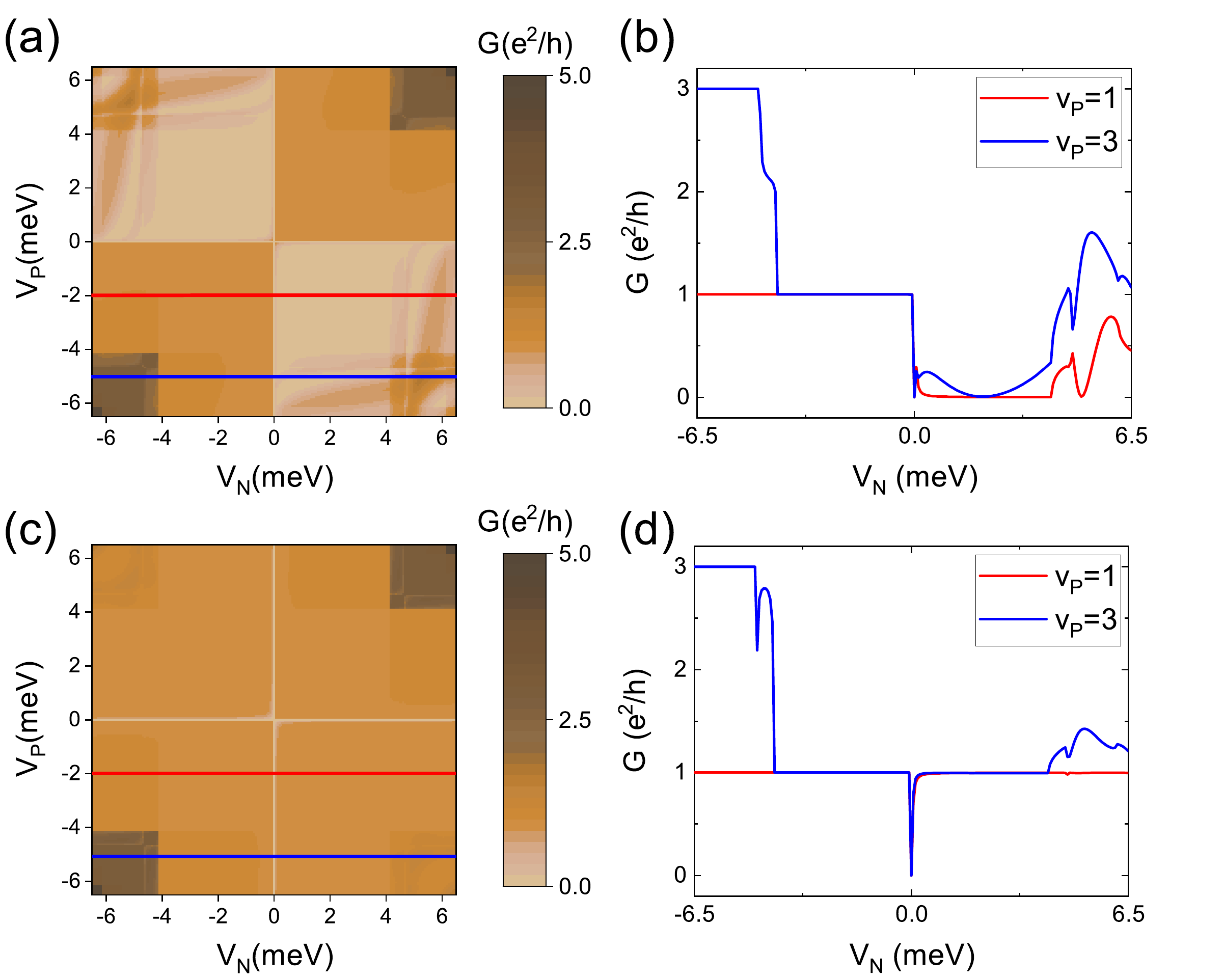}
  \caption{\label{fig31}
  Conductance $G$ in 3D TI PN junction versus the gate voltage $V_P$ and $V_N$ with
  vertical bias $U_T=0$ and $\phi=0$ (a) and $\piup$ (c).
  (b) and (d) show $G$ along the red and blue cut lines in (a) and (c), respectively. The red line depicts $V_P=-2$meV corresponding to $\nu_P=1$, and the blue line depicts $V_P=-5$meV and $\nu_P=3$.
  }
\end{figure}

\begin{figure}
	\includegraphics[width=\linewidth]{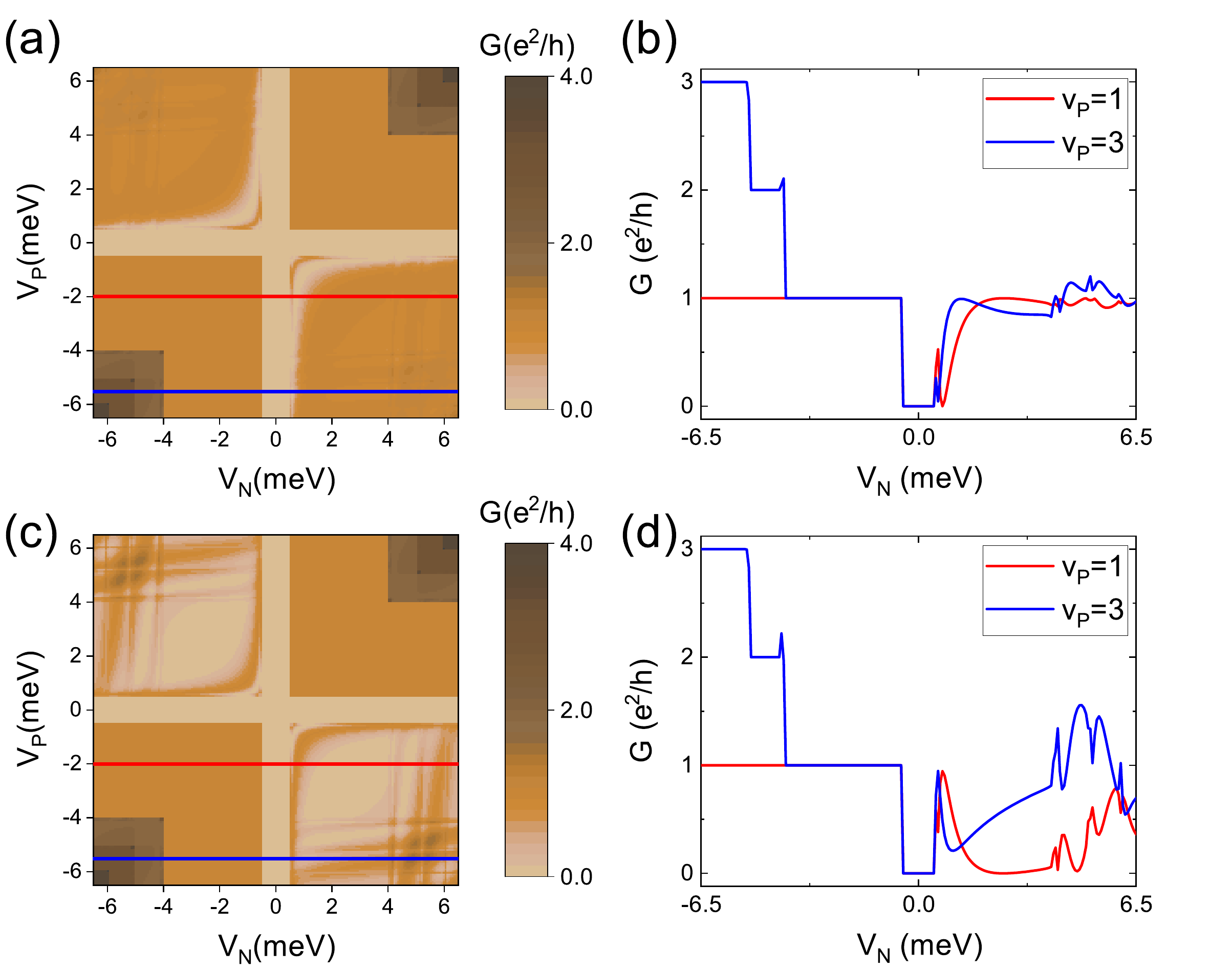}
	\caption{\label{fig32}
  Conductance $G$ in 3D TI PN junction versus the gate voltage $V_P$ and $V_N$ with
  vertical bias $U_T=1$meV and $\phi=0$ (a) and $\piup$ (c).
  (b) and (d) show $G$ along the red and blue cut lines in (a) and (c), respectively. The red line depicts $V_P=-2$meV corresponding to $\nu_P=1$, and the blue line depicts $V_P=-5.5$meV and $\nu_P=3$.}
\end{figure}

In Fig.\ref{fig31} and Fig.\ref{fig32},
we give an overview of the conductance varying with gate voltage $V_P$ and $V_N$
under different axial magnetic flux $\phi$ and vertical bias $U_T$.
The conductance in (a) and (c) of both Fig.\ref{fig31} and Fig.\ref{fig32} is
symmetric about diagonal, i.e. $G(V_P,V_N)=G(V_N,V_P)$.
Because of the current conservation in two-terminal system,
the conductance is invariable when the PN junction rotates $180^\circ$ around z axis,
which changes a $(V_P,V_N)$ junction into $(V_N,V_P)$ junction with $\phi$ changed into $-\phi$,
so $G(V_P,V_N,\phi)=G(V_N,V_P,-\phi)$.
When $\phi=0$, we have $G(V_P,V_N)=G(V_N,V_P)$ as shown in Fig.\ref{fig31}(a) and (c).
In addition, conductance is the periodic function of axial magnetic flux $\phi$
with $G(V_P,V_N,\phi) =G(V_P,V_N,\phi +2\pi)$.
When $\phi=\pi$, $G(V_P,V_N,\pi)=G(V_N,V_P,-\pi) =G(V_N,V_P,\pi)$,
so $G(V_P,V_N)=G(V_N,V_P)$ in Fig.\ref{fig32}(a) and (c).
While the vertical bias $U_T=0$, from Eqs.(\ref{conduct}) and (\ref{thesca}),
one can obtain that the conductance is even function of $\phi$ regardless of other parameters.

In unipolar regime where $V_P$ and $V_N$ hold the same sign,
the conductance presents several plateaus, since the conductance is dependent
on the minimal channel number in the unipolar junction
that $G=\frac{e^2}{h}min(|\nu_P|,|\nu_N|)$.
Specifically, when $U_T=0$ the upper and lower surfaces are symmetric
and the LLs are doubly degenerate [Fig.\ref{fig1}(b)].
Thus the conductance plateaus are odd number in units of $\frac{e^2}{h}$ in Fig.\ref{fig31}.
In Fig.\ref{fig32} as $U_T$ breaking the degeneracy of each LL [Fig.\ref{fig1}(c)],
the conductance plateau can occur at even number.
In particular, there is a large light color cross with $G=0$
in the center of Fig.\ref{fig32}(a) and (c),
referring to the band gap induced by $U_T$ in Fig.\ref{fig1}(c).
In addition, from Fig.\ref{fig31} and Fig.\ref{fig32}, one can see that
the conductances at the first and third quadrant in panel (a)
is almost the same as one in panel (c).
This indicates little effect of the flux $\phi$ in unipolar regime.

On the other hand, in bipolar regime where $V_P$ and $V_N$ have the opposite sign,
the conductance $G$ is strongly dependent on axial magnetic flux $\phi$ and vertical bias $U_T$.
From Fig.\ref{fig31} and Fig.\ref{fig32},
the significant difference between (a) and (c) in the second and fourth quadrant
shows the influence of $\phi$ in bipolar regime.
When $U_T=0$ where upper and lower surfaces are symmetric,
the conductance $G$ is quite small at $\phi=0$ [see Fig.\ref{fig31}(a) and (b)],
but it has the large value at $\phi=\pi$ [Fig.\ref{fig31}(c) and (d)].
At $\phi=\pi$, the conductance presents a plateau with the plateau value $e^2/h$
in $(\nu_P,\nu_N)=(1,-1)$ and $(1,-3)$ regimes [see Fig.\ref{fig31}(d)].
While with a finite $U_T$ where the symmetry between the upper and lower surfaces is broken,
the plateau disappears [see Fig.\ref{fig32}(b) and (d)].
Now $G$ can be large at $\phi=0$ and is small at $\phi=\pi$.

\section{\label{parity} symmetric 3D TI PN junction}

\begin{figure}
	\includegraphics[width=\linewidth]{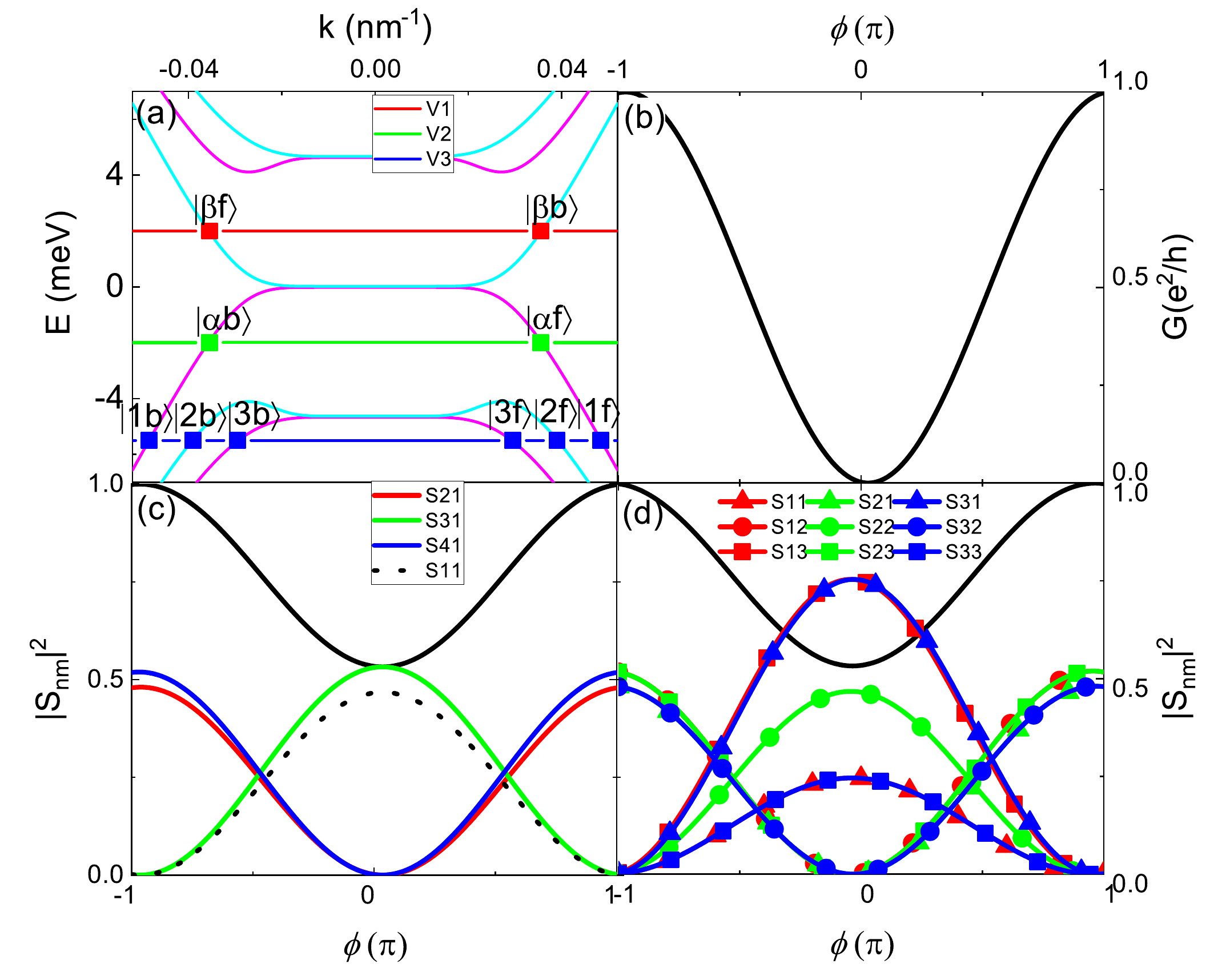}
	\caption{\label{fig4}
(a) The lowest three LLs in Fig.\ref{fig1}(b) with the parity of their branches labeled
by pink ($\alpha=+1$) and cyan ($\alpha=-1$).
Three candidate energies $E=2$meV, $-2$meV and $-5.5$meV are marked corresponding to
filling factors $\nu=1$, $-1$ and $-3$ respectively.
(b) Conductance $G$ versus $\phi$ in (1,-1) PN junction.
(c) The three colored lines depict transmission coefficients $|S_{n1}|^2$ in (1,-3) PN junction,
and the dashed line depicts the reflection coefficient $|S_{11}|^2$
while black line depicts the conductance $G$ in units of $e^2/h$.
(d) The nine colored lines depict the reflection coefficients $|S_{nm}|^2$
in (-3,1) PN junction. The color of each line denotes the outgoing index
and the symbol denotes the incident index.
The black line denotes the conductance $G$ in units of $e^2/h$.}
\end{figure}

In this section we investigate the pristine 3D TI PN junction without vertical bias and disorder.
The influence of the axial magnetic flux $\phi$ in bipolar regime is investigated in Fig.\ref{fig4},
where we numerically simulate the scattering coefficient $|S_{nm}|^2$ by
the Green's function method.
The mirror reflection symmetry about x-y plane is preserved
and we can define the parity $\alpha_n=\pm1$ for longitudinal modes
under the reflection operator $\mathcal{P}$,
\begin{equation}
\mathcal{P}|in(out)\rangle_n=\alpha_n|in(out)\rangle_n.
\end{equation}
The parity of each branch of bands is distinguished by color in Fig.\ref{fig4}(a).

In the appendix, we show that the scattering coefficient $|S_{nm}|^2$ is in the form of
\begin{equation}
|S_{nm}|^2=2|A_{nm}|^2(1+\cos(\phi+\alpha_{nm}\piup)),\label{thesca}
\end{equation}
where $\alpha_{nm}=\frac{1-\alpha_n\alpha_m}{2}$.
Eq.(\ref{thesca}) indicates that all the scattering coefficients harmonically oscillate
with the flux $\phi$, and the initial phase is either 0 or $\piup$,
dependent on the parity factor $\alpha_{nm}$. If the incoming mode $|\mathrm{in}\rangle_m$
and outgoing mode $|\mathrm{out}\rangle_n$ hold the same parity,
the initial phase is 0 and $|S_{nm}|^2$ arrives at the peak at $\phi=0$.
On the other hand, if $|\mathrm{in}\rangle_m$ and $|\mathrm{out}\rangle_n$ hold different parities,
the initial phase is $\piup$ and $|S_{nm}|^2=0$ at $\phi=0$.

The scattering coefficients $|S_{nm}|^2$
are divided into two parts by the outgoing index $n$:
those $n\leqslant\nu_P$ are reflection coefficients and $n>\nu_P$ are transmission coefficients.
The conductance $G$ is the sum of all the transmission coefficients [see Eq.(\ref{conduct})].
The three candidate energy $E=-5.5$meV, $-2$meV and $2$meV [see Fig.\ref{fig4}(a)]
are corresponding to three different filling factor $\nu=-3$, $-1$ and $1$.
In the rest of this section, we repeatedly choose two of these candidate energies
by choosing suitable $V_P$ and $V_N$,
to form various $(\nu_P,\nu_N)$ TI PN junctions
and investigate the scattering coefficients $|S_{nm}|^2$ varying with flux $\phi$.

The $(1,-1)$ PN junction is the most discussed case
since it is in the lowest energy regime.
There is only one transmission coefficient $|S_{21}|^2$
which is exactly proportional to the conductance.
Fig.\ref{fig4}(b) depicts the conductance $G=\frac{e^2}{h}|\langle \mathrm{out}_2|S|\mathrm{in}_1\rangle|^2$.
The $|\mathrm{in}\rangle_1$ is just the $|\beta f\rangle$ in Fig.\ref{fig4}(a)
and $|\mathrm{out}\rangle_2$ is $|\alpha f\rangle$.
These two modes hold different parities. Thus $\alpha_{21}=1$ and the initial phase is $\piup$,
leading that $G=(e^2/2h)(1-\cos\phi)$ and it is $0$ at $\phi=0$ and $e^2/h$ at $\phi=\pi$
[see Fig.\ref{fig4}(b)].
This also explains the $G=0$ and $G=e^2/h$ conductance plateaus in (1,-1) regime
in Fig.\ref{fig31}(b) and Fig.\ref{fig31}(d).

In the $(1,-3)$ PN junction, there are 3 forward outgoing modes
$|\mathrm{out}\rangle_2$, $|\mathrm{out}\rangle_3$ and $|\mathrm{out}\rangle_4$,
which are the state $|1f\rangle$, $|2f\rangle$ and $|3f\rangle$ in Fig.\ref{fig4}(a),
with one incoming mode $|\mathrm{in}\rangle_1$, which is just the $|\beta f\rangle$,
and one backward outgoing mode $|\mathrm{out}\rangle_1$, which is just the $|\beta b\rangle$.
The parity of each mode gives $\alpha_{21}=\alpha_{41}=1$ and $\alpha_{11}=\alpha_{31}=0$.
Thus the initial phase are $\piup$ of $|S_{21}|^2$ and $|S_{41}|^2$
but 0 of $|S_{31}|^2$ and $|S_{11}|^2$.
Fig.\ref{fig4}(c) shows the scattering coefficients $|S_{nm}|^2$,
in which four $|S_{nm}|^2$ exhibit the harmonic oscillations
with $|S_{21}|^2$ and $|S_{41}|^2$ having the initial phase $\piup$ and
$|S_{31}|^2$ and $|S_{11}|^2$ being the initial phase $0$.
Since the reflection coefficient $|S_{11}|^2=0$ at $\phi=\piup$,
the conductance $G=1-|S_{11}|^2$ is always 1 in units of $e^2/h$.
This accounts for the $G=e^2/h$ plateau in the (1,-3) regime in Fig.\ref{fig31}(d).

The parity analysis also works in the multi-mode injecting case.
In the (-3,1) PN junction there are three incoming modes $|\mathrm{in}\rangle_1,|\mathrm{in}\rangle_2,|\mathrm{in}\rangle_3$
which are the states $|1f\rangle,|2f\rangle,|3f\rangle$ in Fig.\ref{fig4}(a),
along with three backward modes $|\mathrm{out}\rangle_1,|\mathrm{out}\rangle_2,|\mathrm{out}\rangle_3$
which are the states $|1b\rangle,|2b\rangle,|3b\rangle$.
The nine reflection coefficients are depicted in Fig.\ref{fig4}(d),
distinguished by color and symbol of each curve.
The color denotes the outgoing mode while the symbol denotes the incoming mode.
It is clear that $|S_{11}|^2, |S_{22}|^2, |S_{33}|^2,
|S_{13}|^2, |S_{31}|^2$ have the parity coefficient $\alpha_{nm}=0$,
and they have the maximum at $\phi=0$ and are zero at $\phi=\pi$.
The other four reflection coefficients
$|S_{12}|^2, |S_{21}|^2, |S_{23}|^2, |S_{32}|^2$ have $\alpha_{nm}=1$,
leading that they are zero at $\phi=0$ and reach the maximum at $\phi=\pi$ [see Fig.\ref{fig4}(d)].
Moreover, the combination of reflection about x-z plane
and time reversion is conserved in this 3D TI PN junction,
and this exchanges the $|\mathrm{in}\rangle_n$ and $|\mathrm{out}\rangle_n$ for $n=1,2,3$.
For this reason the reflection coefficients have the property $|S_{nm}|^2=|S_{mn}|^2$.
In addition, due to the current conservation in two-terminal system,
the conductance $G$ of the (-3,1) PN junction is exactly equal to that
of the (1,-3) PN junction [see Fig.\ref{fig4}(c) and (d)].

\section{\label{bias} non-symmetric 3D TI PN junction in the nonzero vertical bias}

\begin{figure}
	\includegraphics[width=\linewidth]{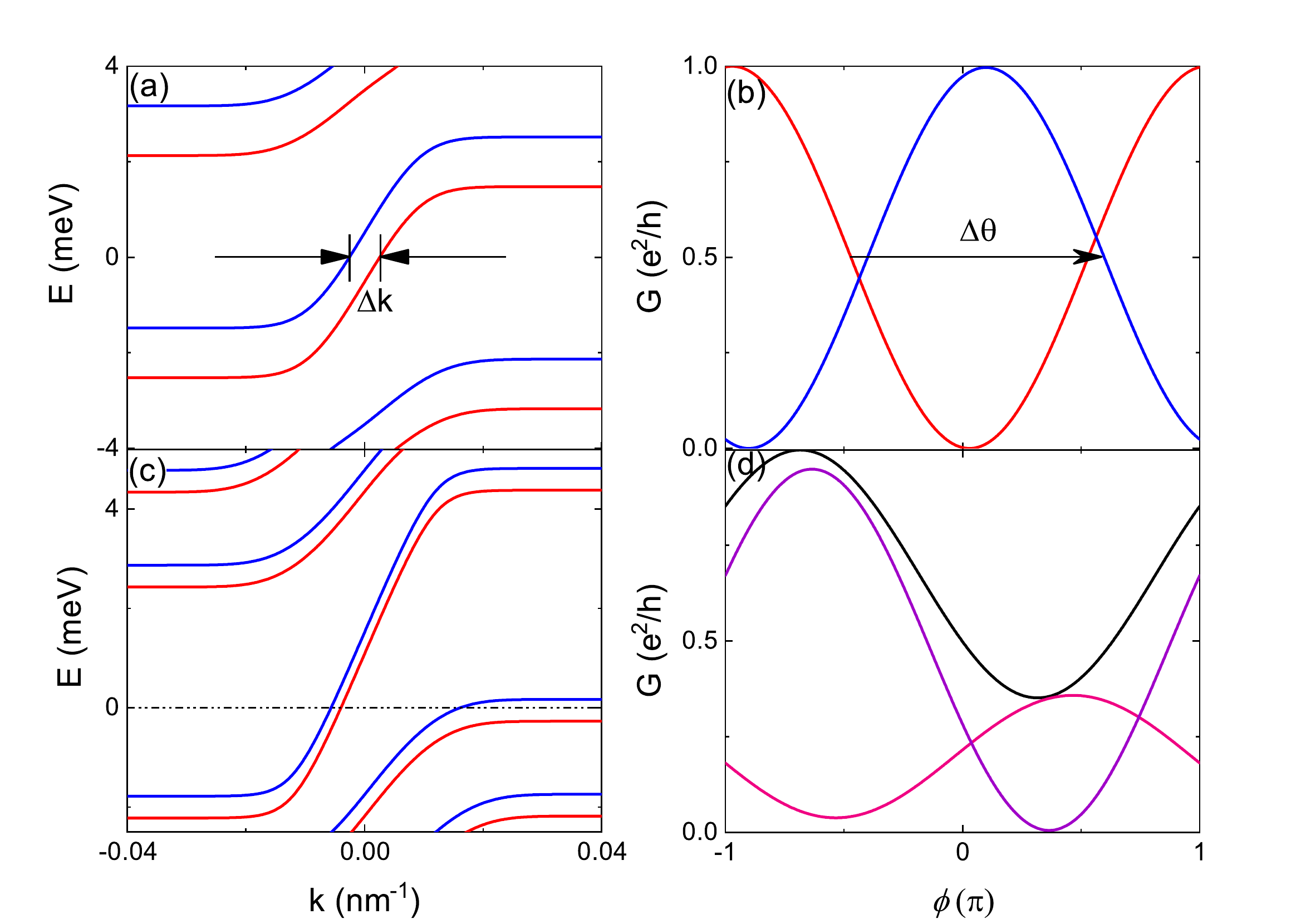}
	\caption{\label{fig5}
(a) and (c) Energy spectrum of transverse modes.
Red and blue curves depict the energy bands of transverse modes in upper and lower surfaces, respectively.
(a) The (1,-1) PN junction with $V_P=-2$meV, $V_N=2$meV and $U_T=1$meV.
(b) A comparison of the conduction $G$ with and without $U_T$.
The blue curve depicts the conductance of the junction in (a).
The red curve depicts the conductance of a junction with the same $V_P$ and $V_N$ but $U_T=0$.
The blue curve has a phase shift $\Delta\theta$ with respect to the red curve.
(c) The (1,-2) PN junction with $V_P=-2$meV, $V_N=4.6$meV and $U_T=0.4$meV.
(d) The transmission coefficients (colored curve) and the conductance (black curve) of
the (1,-2) PN junction in (c).}
\end{figure}

The vertical bias $U_T$ breaks the symmetry of upper and lower surfaces
and breaks the parity as well.
The scattering coefficient $|S_{nm}|^2$ is no more Eq.(\ref{thesca})
but in a more general form (see appendix)
\begin{equation}
|S_{nm}|^2=|A_{nm}|^2+|B_{nm}|^2+2|A_{nm}B_{nm}|\cos(\phi+\theta_{nm}).\label{coef1}
\end{equation}
The initial phase $\theta_{nm}$ is an arbitrary value instead of 0 or $\piup$ only.
This phase shift is due to the momentum difference between transverse modes of upper and lower surfaces, which is called kinetic phase.
In order to investigate these transverse modes, an infinite wide 3D TI PN junction is considered
(i.e. the width $L \rightarrow \infty$), and its energy spectrum is shown in
Fig.\ref{fig5}(a) and (c). From the energy spectrum, the transverse modes can be obtained
from the intersection of the energy band and Fermi level.

The transverse modes come from LLs which distort at the PN interface
and cross the Fermi level.
The vertical bias $U_T$ lifts the energy in upper surface by $U_T/2$,
and $-U_T/2$ in lower surface.
This leads to a momentum difference $\Delta k$ between upper and lower transverse modes [Fig.\ref{fig5}(a)]. In a (1,-1) PN junction, this difference straightforwardly induces a kinetic phase shift $\Delta\theta$
in conductance oscillation [see Fig.\ref{fig5}(b)].
The relation between $\Delta k$ and $\Delta\theta$ is detailed in Sec.\ref{phase}.

In some cases, $U_T$ creates a new crosspoint of LLs and Fermi level at one surface,
inducing an additional transverse mode [Fig.\ref{fig5}(c)].
This results in an even number of filling factor $\nu_N=-2$ in the exit region
for the sake of mode conservation.
For this reason there are two transmission coefficients in total,
which are depicted in Fig.\ref{fig5}(d).
The initial phase of each scattering coefficient is not restricted to 0 or $\piup$,
but an arbitrary value as in Eq.(\ref{coef1}). This (1,-2) PN junction presents a
novel spin precession phenomenon which is illuminated in Sec.\ref{precession}.

\subsection{\label{phase}kinetic phase in (1,-1) TI PN junction}

\begin{figure}
	\includegraphics[width=\linewidth]{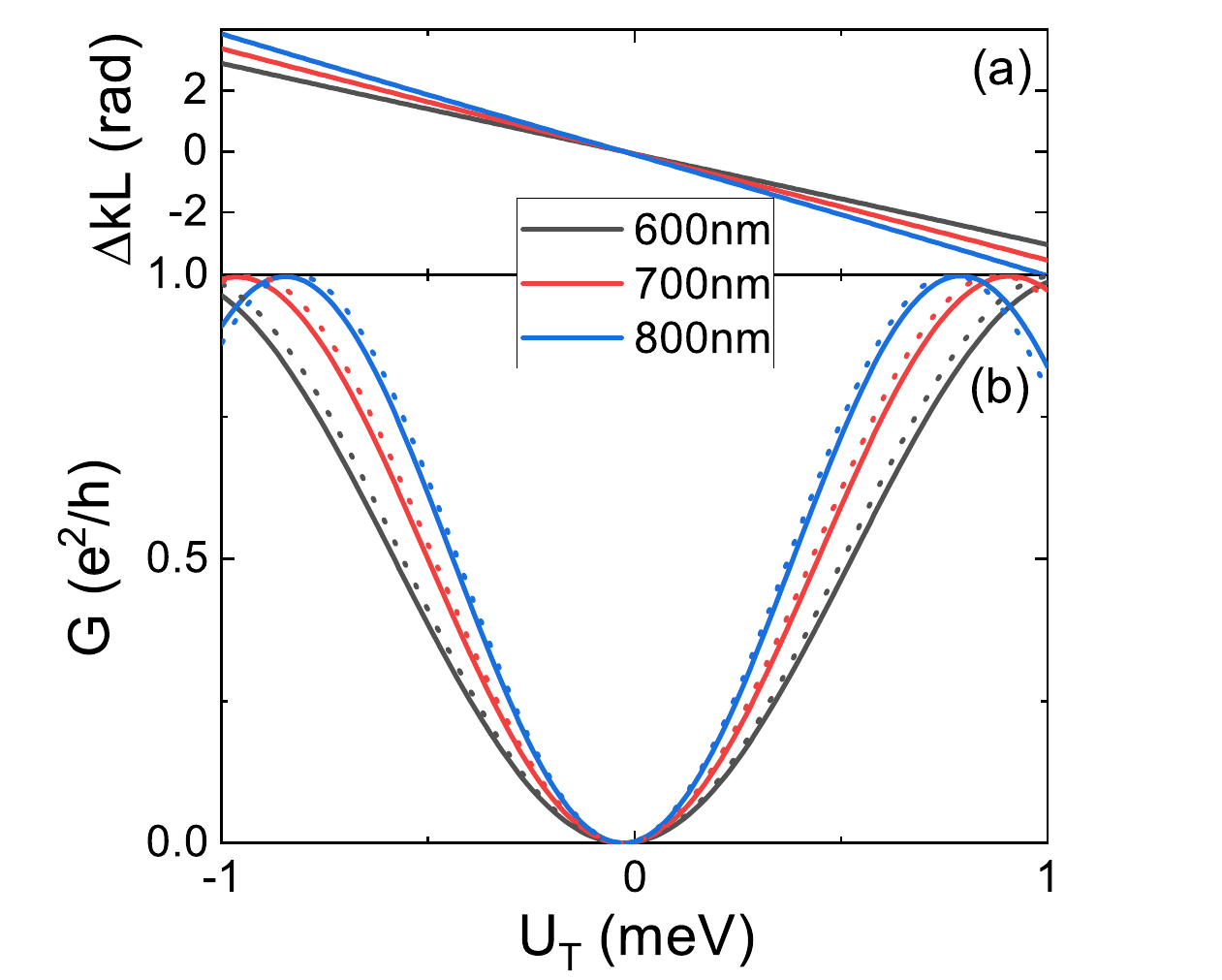}
	\caption{\label{fig6}
(a) Kinetic phase $\Delta k L$ versus vertical bias $U_T$.
(b) Comparing the numerical conductance $G$ (solid line) with the analytical conductance by Eq.(\ref{dedu}) (dotted line). The parameters are $\phi=0$, $V_P=-2$meV, $V_N=2$meV, and
the PN junction width $L=600$, $700$ and $800$nm.
}
\end{figure}

The kinetic phase shift in Fig.\ref{fig5}(b) can be derived from Eq.(\ref{coef1}).
It gives the conductance $G$ in (1,-1) TI PN junction (see Appendix):
\begin{equation}
G=\frac{e^2}{2h}[1-\cos(\phi+\Delta kL)].\label{dedu}
\end{equation}
This deduction consists with our numerical results, see Fig.\ref{fig6}.
With the help of the infinite wide TI PN junction,
the transverse modes and the momentum difference $\Delta k$
can be solved out directly by the transfer matrix method.\cite{my}
Then we can get the kinetic phase $\Delta kL$ of every $U_T$,
as well as the analytical conductance $G$ in Eq.(\ref{dedu}).
Meanwhile, the conductance $G$ can numerically calculated by the Green's function method.
Fig.\ref{fig6}(b) shows both analytical and numerical conductances
for three different size of (1,-1) TI PN junctions as a comparison.
Since $U_T$ is small compared with the scale of the energy band,
it changes $\Delta k$ linearly.
The kinetic phase $\Delta kL$ varies proportionally to $U_T$ [see Fig.\ref{fig6}(a)],
causing a harmonic oscillation in conductance $G$.
Although the analytical $G$ is close enough to the numerical $G$,
there exists a small deviation, indicating the indeed phase shift $\Delta\theta$
is slightly lesser than the ideal kinetic phase $\Delta kL$.
This might be attributed to the boundary effect,
that the distance for transverse mode going through is not the full length of $L$
but the majority of it.
This explanation also fits the fact that the deviation is small in long $L$ but large in short $L$.

\subsection{\label{precession} spin precession in (1,-2) TI PN junction}

\begin{figure}
	\includegraphics[width=\linewidth]{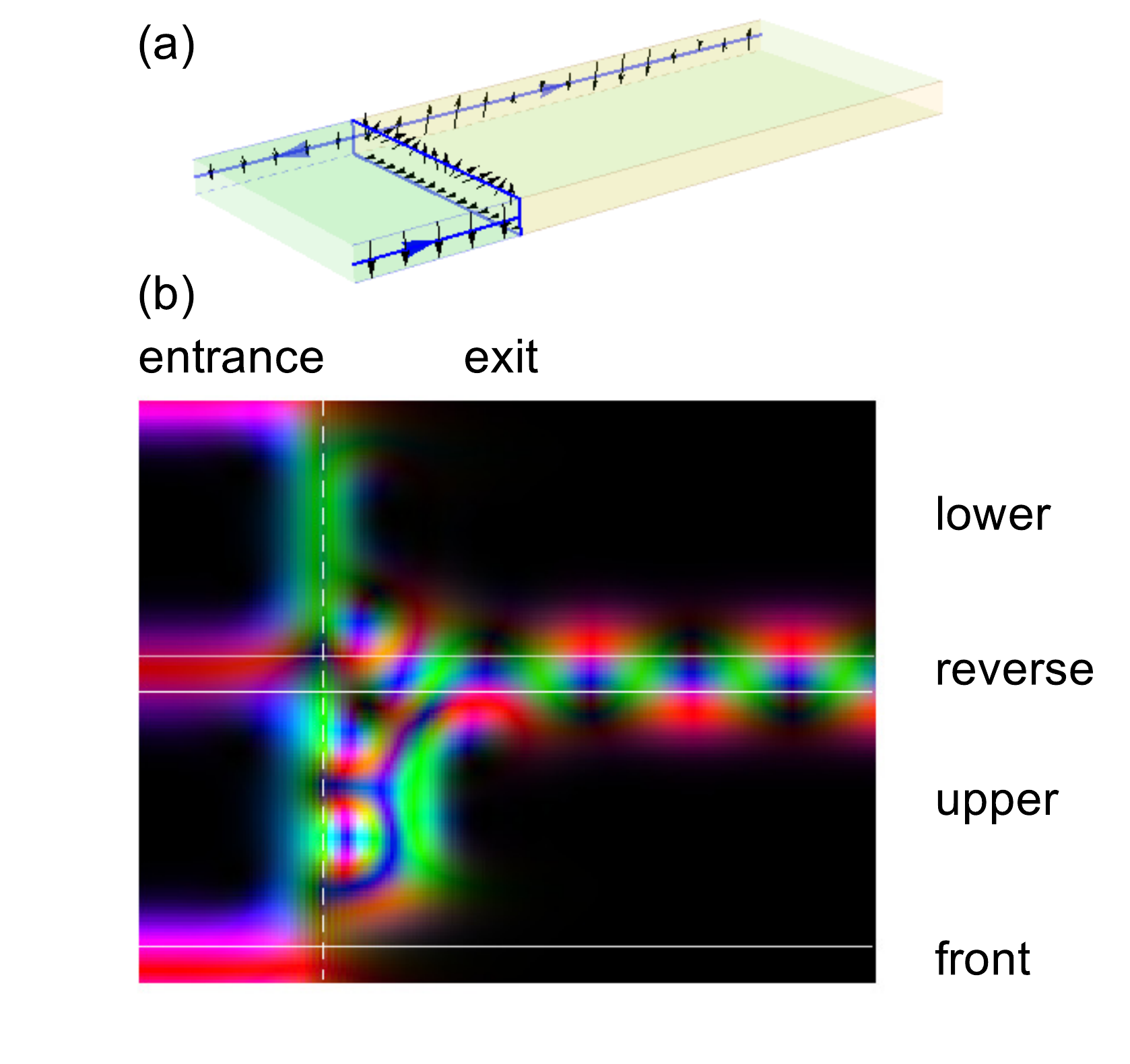}
	\caption{\label{fig7}
(a) A sketch of spin precession.
Black arrows represent the spin polarization and blue arrow represents the current.
Spin aligns at the entrance region but rotates at the exit region.
(b) A RGB color map of the spin distribution in the surface of (1,-2) PN junction.
The parameters are consistent with in Fig.\ref{fig5}(c) and $\phi=0$.
The red, green, blue represent three different polarization of tangential, axial, normal, respectively.
This is expanded view and the white solid lines represent the edges of 3D TI
while dashed line represents the PN interface.}
\end{figure}

When the vertical bias $U_T=0$, the (1,-2) TI PN junction is forbidden
since the LLs are doubly degenerate and the filling factors are odd number.
With the help of $U_T$, (1,-2) TI PN junction is achieved
and a novel spin precession appears.
The spin polarization aligns along the side surfaces in the entrance region,
but rotates along the reverse side surface
in the exit region [see Fig.\ref{fig7}(a)].
For the (1,-2) 3D TI PN junction, there exists
an incoming mode $|\mathrm{in}\rangle_1$ on the front side surface,
a reflecting mode $|\mathrm{out}\rangle_1$ and
two transmitting modes $|\mathrm{out}\rangle_2$ and $|\mathrm{out}\rangle_3$
on the reverse side surface,
and three transverse modes $|\mathrm{low}\rangle_1$, $|\mathrm{up}\rangle_1$ and
$|\mathrm{up}\rangle_2$ on the PN interface.
The spin precession phenomenon is due to the interference of $|\mathrm{out}\rangle_2$ and $|\mathrm{out}\rangle_3$, and it can be used in the spin valve device\cite{svalve}.

In order to illustrate the spin precession phenomenon,
we simulate a sample of (1,-2) 3D TI PN junction of which
the parameters are consistent with Fig.\ref{fig5}(c),
showing its spin distribution by expanded view in Fig.\ref{fig7}(b).
Spin polarization on the surface of 3D TI PN junction is described with the basis
$(s_\parallel,s_x,s_\perp)$, with $s_x$ being along the x axis,
$s_\perp$ being the normal direction of the surface, and
$s_\parallel=s_x\times s_\perp$ being the tangential direction.
The spin polarization is fixed on the front side and reverse side surfaces
in the entrance region, as well along the transverse mode on the lower surface,
since these areas contain only a single transmission mode.
The wavefunction changes only at the overall phase factor by $\me^{\mi kx}$,
which does not effect the value and direction of spin.
In Fig.\ref{fig7}(b), these areas with fixed polarization
are presented monotonous in color.
On the other hand, there presents complicated structures periodically arranged
along the PN interface on the upper surface and on the reverse side surface
in the exit region.
These two areas are occupied by two different modes
which interfere with each other producing a beat pattern,
in which the spin precession occurs, i.e. the electron's spin rotates
as the electron moves ahead [see Fig.\ref{fig7}(a)].
The spin precession originates from the two transmitting modes having different momentums.
The beat in the exit region appears only at $|\nu_N|=2$,
because more than two different frequencies mixed together would break the periodical beat pattern.

\begin{figure}
\includegraphics[width=\linewidth]{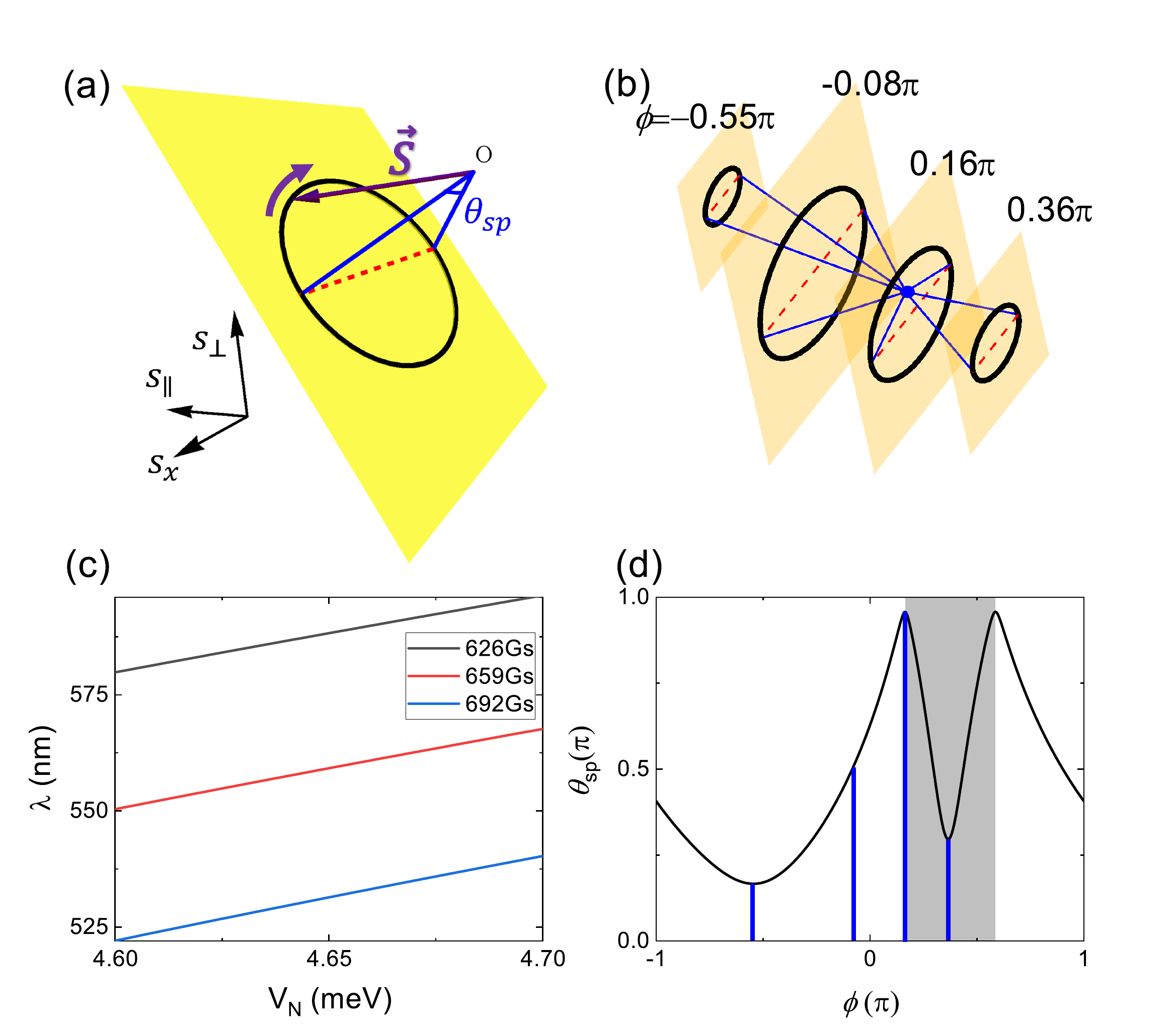}
\caption{\label{fig8}
(a) With the electron moving forward in the reverse side in exit region,
its spin polarization vector $\vec{S}$ moves along the elliptic orbit in the spin space.
The longer axis of the ellipse (red dashed line) has an open angle $\theta_{sp}$ to the ordinate origin, which is used to denote the degree of spin variation.
(b) The orbit varies in position and size with the change of magnetic flux $\phi$.
The blue lines depict open angles of longer axis.
(c) The space period $\lambda$ of the spin precession
in the exit region varying with the gate voltage $V_N$ under different magnetic field $B_\perp$.
(d) The opening angle $\theta_{sp}$ versus the axial magnetic flux $\phi$. The parameters
are the same as Fig.\ref{fig5}(c). The four blue vertical lines correspond to the four orbits in (b). The light (dark) background denotes that the orbital plane is at one side (the other side) of the ordinate origin. When the orbital plane passes the ordinate origin, $\theta_{sp}$ reaches its peak.
}
\end{figure}

The wavefunction in the exit region is
\begin{equation}
|\Psi\rangle=S_{21}|\mathrm{out}\rangle_2+S_{31}|\mathrm{out}\rangle_3\label{Psi}.
\end{equation}
As shown in Fig.\ref{fig5}(d), the transmission coefficients $|S_{21}|^2$
and $|S_{31}|^2$ vary with the axial magnetic flux $\phi$.
However, the space period $\lambda$ of the spin precession is unrelated to $\phi$.
It is decided directly by the momentums of the two transmitting modes, $k_2$ and $k_3$,
\begin{equation}
\lambda=\frac{2\piup}{|k_2-k_3|}.
\end{equation}
These momentums are dependent on the energy spectrum of exit region
which can be tuned by gate voltage $V_N$ and the magnetic field $B_\perp$.
Fig.\ref{fig8}(c) shows that the space period $\lambda$ of the spin precession increases
with the gate voltage $V_N$ and decreases with the magnetic field $B_\perp$.

Considering an electron moves ahead in the reverse side in exit region,
its spin polarization vector rotates along a trajectory in the spin space
of $(s_\parallel,s_x,s_\perp)$ [See Fig.\ref{fig8}(a)].
This trajectory is a closed orbit because of the periodicity.
Because of spin-momentum locking, $\langle out_i |\hat{s}_x|out\rangle_i=0$ for $i=2,3$,
this means that the axial spin component is zero
for every single longitudinal mode.\cite{core}
But the cross term $\langle out_2|\hat{s}_x| out\rangle_3 \neq 0$.
By using the wavefunction in Eq.(\ref{Psi}), the spin polarization vector $\vec{S}$ can be obtained
that the orbit is always an ellipse of which the longer axis is parallel to $s_x$.
We introduce an opening angle $\theta_{sp}$ of this longer axis to the ordinate origin
to denote the changes of the spin direction
in a spin precession period. This opening angle is influenced by magnetic flux $\phi$ [See Fig.\ref{fig8}(b)] and is numerically analyzed in Fig.\ref{fig8}(d).
The flux $\phi$ goes through a cycle from $-\piup$ to $\piup$,
and the plane of orbit passes the ordinate origin twice in the cycle,
forming two peaks in curve $\theta_{sp}$ versus $\phi$ in Fig.\ref{fig8}(d).
At the peaks, $\theta_{sp}$ can almost reach the largest value $\pi$.
With the change of the system's parameters,
these two peaks with the large $\theta_{sp}$ can usually exist.
Notice that a large opening angle $\theta_{sp}$ in the spin precession
is very conducive to spin valve devices.

\section{\label{disorder} effect of disorder on the transport through TI PN junction}

In practice, the symmetry of upper and lower surfaces is often broken even without vertical bias.
Since the 3D TI is always placed on a substrate\cite{substrate1,substrate2},
the lower surface is imperfect while the upper is relatively perfect.
In this section we focus on a (1,-1) 3D TI PN junction
of which the lower surface is disordered, and the vertical bias $U_T=0$.
This is the simplest case since there is only one single mode in every part of the PN junction.
The space distribution indicates that only the transverse mode $|\mathrm{low}\rangle_1$
in the lower surface is effected by the disorder.
For every single disorder configuration, the mode $|\mathrm{low}\rangle_1$
gets a certain extra phase factor $\me^{\mi\theta}$
after going though the disorder region.
The conductance for this configuration is
\begin{equation}
G(\phi ,\theta)=\frac{e^2}{2h}(1-\cos(\phi+\theta)).
\end{equation}
The factor $\theta$ is random with the disorder configuration.
However, given a group of certain parameter for disorder,
the phase $\theta$ obeys a normal distribution $f(\theta)$
for the reason of central limit theory.
It gives the expectation of conductance for such a group of parameter:
\begin{equation}
G(\phi)=\frac{e^2}{2h}\int f(\theta)(1-\cos(\phi+\theta))\dif\theta\label{rerw} .
\end{equation}
Using a intensity factor $F$ and a phase factor $\varphi$ to describe the distribution $f(\theta)$
\begin{equation}
F\me^{\mi\varphi}=\int\me^{\mi\theta}f(\theta)\dif\theta ,
\end{equation}
Eq.(\ref{rerw}) is rewritten as
\begin{equation}
G(\phi)=\frac{e^2}{2h}(1-F\cos(\phi+\varphi))\label{discon} .
\end{equation}

In order to investigate the behavior of the factor $F$ and $\varphi$ with parameters of disorder,
we simulate a sample with its length $L=2\mu$m and the magnetic field $B_\perp=132$Gs.
Space correlated disorder is expected to be distributed on the whole lower surface.
Here in the simulation, the disorder is added only near the PN interface
in the middle of the lower surface.
The disorder has two different effects on the transport of 3D TI PN junction,
that the phase shift $\theta$ for $|\mathrm{low}\rangle_1$
and the back scattering from $|\mathrm{in}\rangle_1$ to $|\mathrm{out}\rangle_1$.
By selectively disordering near the PN interface, we can
focus on the extraordinary phase shift $\theta$ rather
than the ordinary back scattering of chiral edge states at the opposite sides.
In real device, the back scattering at the opposite sides is very weak
due to the large size and the strong magnetic field $B_\perp$.
In the disorder region, the on-site energy is added by a disorder term:\cite{disorderadd1}
\begin{equation}
\Omega_i=\frac{\sum\limits_j\tilde{\Omega}_j\exp(-|\vec{r}_{ij}|^2/2\eta^2)}
 {\sqrt{\sum\limits_j\exp(-|\vec{r}_{ij}|^2/\eta^2)}} .\label{dis1}
\end{equation}
Here $\eta$ is the correlation length.
$\Omega_i$ represents the on-site energy fluctuation on site $i$
while $\tilde{\Omega}_j$ represents the strength of the disorder spot of which the center is on site $j$.
In the disorder region, the density of the disorder spot is $\rho$ and
$\tilde{\Omega}_j$ are uniformly distributed in the range of [$-0.5w,0.5w$],
where $w$ is the disorder strength.
So the disorder configuration is described by three parameters:
correlation length $\eta$, disorder strength $w$, and disorder density $\rho$.

\begin{figure}
	\includegraphics[width=\linewidth]{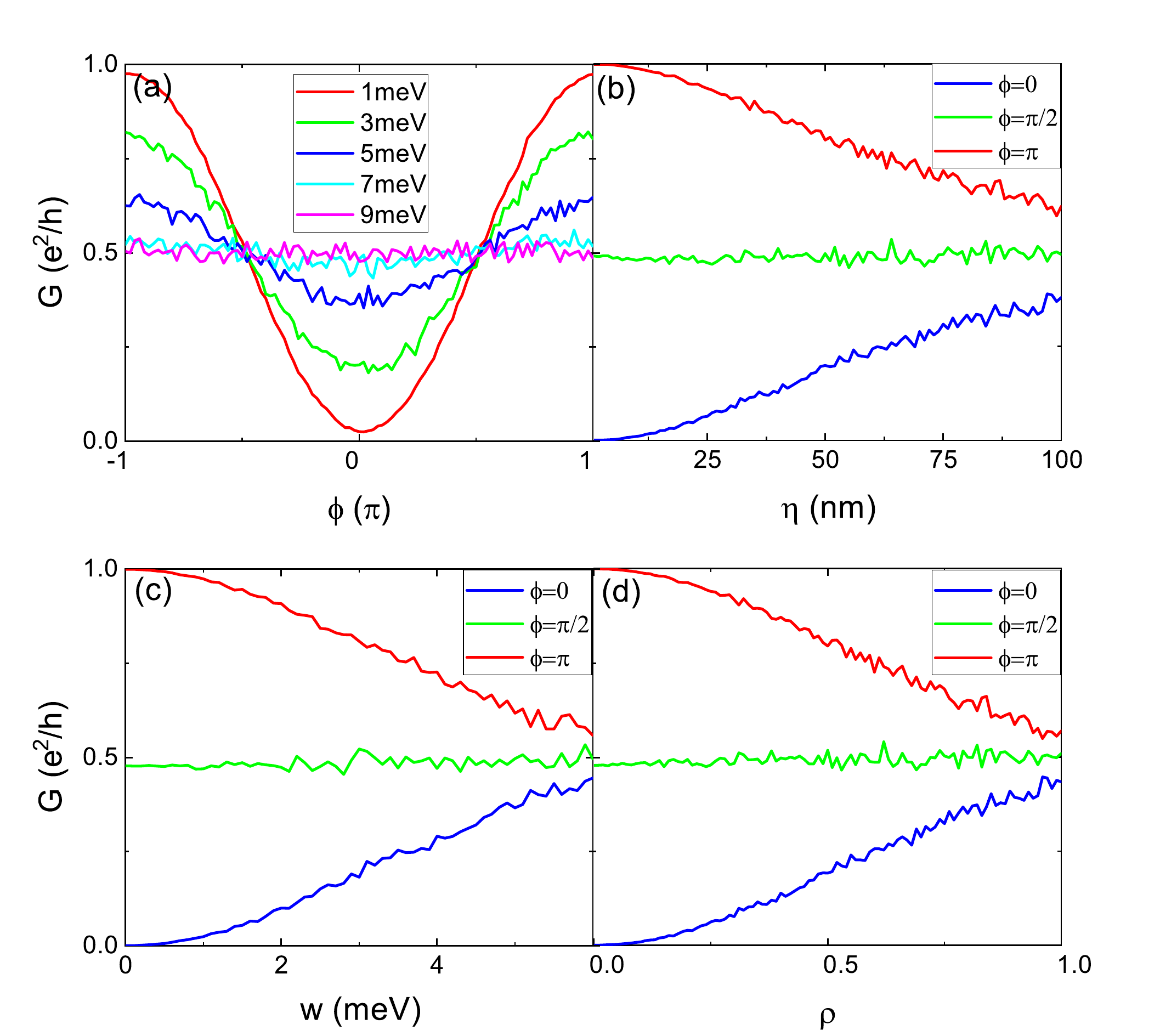}
	\caption{\label{fig9}
Numerical simulation for disordered 3D TI PN junction with
$V_P=-1$meV, $V_N=1$meV and $U_T=0$.
Except as otherwise noted, the default parameters are $\eta=50$nm, $\rho=1$.
(a) Conductance $G$ varying with the flux $\phi$ under different disorder strength $w$.
(b) $G$ varying with the correlation length $\eta$ under different flux $\phi$
with disorder strength $w=3$meV.
(c) $G$ varying with the disorder strength $w$ under different flux $\phi$.
(d) $G$ varying with the disorder density $\rho$ under different flux $\phi$
with disorder strength $w=6$meV.}
\end{figure}

The numerical simulation of the conductance $G$ in disordered (1,-1) 3D TI PN junction
is shown in Fig.\ref{fig9}.
Fig.\ref{fig9}(a) illustrates $G$ oscillating with $\phi$ under different disorder strength $w$,
and Fig(b-d) depict the conductance $G$ at different flux $\phi$ varying with correlation length $\eta$,
disorder strength $w$ and disorder density $\rho$, respectively.
Each data point in Fig.\ref{fig9} is an average of 500 results of different disorder configurations. In fact, the average among a large ensemble of disorder configuration
is equivalent to consider the dephasing effect.\cite{ar3}

Fig.\ref{fig9} reveals the undetermined coefficients $\varphi$ and $F$ in Eq.(\ref{discon}).
The averaged phase shift is always $\varphi=0$
because all the curves in Fig.\ref{fig9}(a) arrive their peaks at $\phi=\pm\piup$,
arrive their valleys at $\phi=0$ and converge at $\phi=\pm\piup/2$ with $G=0.5e^2/h$.
Moreover, $\varphi=0$ is also illustrated in Fig.\ref{fig9}(b-d) since the conductance presents
as a plateau at $G=0.5e^2/h$ in all these figures at $\phi=\piup/2$,
pointing to $\varphi=0$ in Eq.(\ref{discon}).

As the phase shift $\varphi$ is recognized to be 0 unrelated with any parameter,
Eq.(\ref{discon}) gives $G=\frac{e^2}{2h}(1-F)$ at $\phi=0$
and $G=\frac{e^2}{2h}(1+F)$ at $\phi=\piup$,
corresponding to the blue and red line in Fig.\ref{fig9}(b-d).
These figures indicate that the intensity factor $F$ decreases monotonously
with all the three parameters: correlation length $\eta$, disorder strength $w$, and disorder density $\rho$.
For a large correlation length $\eta$,
the random disorder energy $\Omega_i$ changes slowly with the position $i$
at a specific disorder configuration which increases
the effective action of the disorder,\cite{disorderadd1} leading that $F$ is small at the large $\eta$.
The decrease of factor $F$ leads to a suppression of amplitude in $G$ vs $\phi$.
As $F$ approaches to 0, the $G\sim\phi$ curve approaches to a plateau at $0.5e^2/h$ [Fig.\ref{fig9}(a)].
This result is similar with one in the graphene PN junction.\cite{ascience1,ascience2,ar4,ar5}

\section{\label{conclusion}Conclusion}

In summary, we have thoroughly investigated the electron transport through the
3D topological insulator PN junction under the perpendicular and axial magnetic fields.
The transport properties are described by the scattering coefficients
which can be calculated by using the Green's function method.
When the vertical bias and disorder are both absent,
the mirror symmetry about x-y plane is conserved and the upper and lower surfaces are symmetric.
In this case the longitudinal incoming and outgoing modes hold a definite parity of even or odd.
The scattering coefficients harmonically oscillates with the axial magnetic flux
and the oscillation initial phase is $0$ or $\pi$ which depends
on the parities of the incoming and outgoing modes.
The vertical bias breaks the upper-lower surface symmetry and induces an potential difference
between upper and lower surfaces,
which generates a difference in the momentum of the transverse modes
and causes a non-zero kinetic phase.
The kinetic phase leads to a phase shift in the oscillation of
the scattering coefficient versus axial magnetic flux.
In some cases, the vertical bias changes the number of modes and
generates a special (1,-2) 3D topological insulator PN junction,
in which a novel spin precession phenomenon happens.
Here the spin vector rotates on an ellipse orbit while
the electron moving along the side in the exit region.
This spin precession is due to the interference of the outgoing modes
and it has potential applications in the spin valve device.
While in the presence of the disordered case, a single disorder configuration
on the lower surface leads to a phase shift in conductance oscillation.
However, this phase shift vanishes when averaged in a large configuration ensemble,
and the averaged conductance oscillation shows a suppression in amplitude.

\section*{Acknowledgments}
We gratefully acknowledge the financial support from
National Key R and D Program of China (Grant No. 2017YFA0303301),
NBRP of China (Grant No. 2015CB921102), NSF-China under Grants No. 11574007, and
the Key Research Program of the Chinese Academy of Sciences (Grant No. XDPB08-4).

\begin{appendix}
\section{the formula of the scattering coefficient}

In this appendix, we develop the expression of scattering coefficient $|S_{nm}|^2$.
Using scattering matrix $S_1$ to describe how incoming states $|\mathrm{in}\rangle$
is scattered into transverse modes $|\mathrm{up}\rangle$ and $|\mathrm{low}\rangle$,
and $S_2$ describe $|\mathrm{up}\rangle$ and $|\mathrm{low}\rangle$ to $|\mathrm{out}\rangle$.
By taking a rotation $180^\circ$ around x axis with combining a time inversion
transformation, the scattering matrix $S_1$ changes into $S_2$,
so we get a symmetry of
\begin{equation}
\begin{array}{c}
\langle\mathrm{up}_i|S_1|\mathrm{in}_m\rangle=\langle\mathrm{out}_m|S_2|\mathrm{low}_i\rangle ,\\
\langle\mathrm{low}_j|S_1|\mathrm{in}_m\rangle=\langle\mathrm{out}_m|S_2|\mathrm{up}_j\rangle.
\end{array}\label{sym}
\end{equation}

The scattering from $|\mathrm{in}\rangle$ to $|\mathrm{out}\rangle$ is
a combination of $S_1$ and $S_2$, along with a phase variation $\Phi_i = k_i L +\phi/2$
($\Phi_j = k_j L -\phi/2$ ) due to the kinetic phase and AB phase in
the transverse modes of the upper surface (lower surface).
Thus we have
\begin{equation}
S=S_2\Phi S_1 .
\end{equation}
Introducing two complex number $A_{nm}$ and $B_{nm}$ to evaluate the contribution
of upper and lower surfaces, respectively,
\begin{equation}
\begin{array}{l}
A_{nm}=\sum\me^{\mi k_iL} \langle out_n|S_2|\mathrm{up}\rangle_i \langle up_i|S_1|\mathrm{in}\rangle_m ,\\
B_{nm}=\sum\me^{\mi k_jL} \langle out_n|S_2|\mathrm{low}\rangle_j \langle low_j|S_1|\mathrm{in}\rangle_m .
\end{array}
\end{equation}
Using $\theta_{nm}$ to denote the phase difference between $A_{nm}$ and $B_{nm}$
\begin{equation}
\theta_{nm}=\mathrm{rad}(A_{nm})-\mathrm{rad}(B_{nm}).
\end{equation}
Then we have a harmonic oscillation expression for $|S_{nm}|^2$
\begin{eqnarray}
|S_{nm}|^2 & = &|A_{nm} e^{i\phi/2} +B_{nm}e^{-i\phi/2}|^2 \label{coef} \\
& = & |A_{nm}|^2+|B_{nm}|^2+2|A_{nm}B_{nm}|\cos(\phi+\theta_{nm}). \nonumber
\end{eqnarray}
The Eq.(\ref{coef}) gives the most general expression of scattering coefficient
which is valid whether the vertical bias $U_T$ is applied or not.

When $U_T=0$, the horizontal mirror reflection $\mathcal{P}$ preserved and
the upper and lower surfaces are symmetric, so
\begin{equation}
\mathcal{P}|\mathrm{up}\rangle_i=|\mathrm{low}\rangle_i .
\end{equation}
The longitudinal modes $|\mathrm{in}\rangle$ and $|\mathrm{out}\rangle$ hold
definite parity of $\mathcal{P}$, that
\begin{equation}
\mathcal{P}|\mathrm{in(out)}\rangle_n=\alpha_n|\mathrm{in(out)}\rangle_n ,
\end{equation}
where $\alpha_n=\pm1$. Thus we have
\begin{equation}
A_{nm}=(-1)^{\alpha_m\alpha_n}B_{nm}\label{pari}
\end{equation}
and Eq.(\ref{coef}) is turned into
\begin{equation}
|S_{nm}|^2=2|A_{nm}|^2(1+\alpha_m\alpha_n\cos\phi) ,
\end{equation}
which is the Eq.(\ref{thesca}) in Sec.\ref{parity}.

As an simplest case in $(\nu_P,\nu_N)=(1,-1)$ PN junction,
where exists only one single mode in the upper and lower surfaces respectively,
it gives
\begin{equation}
\langle\mathrm{up}_1|S_1|\mathrm{in}_1\rangle=\langle\mathrm{low}_1|S_1|\mathrm{in}_1\rangle=1/\sqrt{2}
\end{equation}
for the sake of spin matching at the vertex.\cite{core}
Combined with Eq.(\ref{sym}), this lead to
\begin{equation}
|A_{11}|=|B_{11}|=1/2 .
\end{equation}
Combined with current conservation, we get
\begin{equation}
|A_{12}|=|B_{12}|=1/2.
\end{equation}
At $U_T=0$, $|\mathrm{in}\rangle_1$ and $|\mathrm{out}\rangle_2$ are at different branches of the 0th LL,
and hold opposite parities. Eq.(\ref{pari}) gives
$A_{12}=-B_{12}$ and $|S_{nm}|^2=\frac{1}{2}(1-\cos\phi)$.
While with a finite vertical bias $U_T$ but still in the $(1,-1)$ PN junction regime,
we have $\theta_{12}=\piup+\Delta kL$ in Eq.(\ref{coef}).
Here $\Delta k=k_{up}-k_{low}$ is the momentum difference between the upper and lower
transverse modes. Then the conductance is
\begin{equation}
G=\frac{e^2}{2h} [1-\cos(\phi+\Delta kL)].
\end{equation}

\end{appendix}

\end{document}